\documentclass[a4paper,11pt]{article}
\pdfoutput=1 

\usepackage{jinstpub} 
\usepackage{lineno,hyperref}
\usepackage[utf8]{inputenc} 
\usepackage{longtable}
\usepackage{verbatim}
\usepackage{textcomp}
\usepackage{wasysym}
\usepackage{eurosym}
\usepackage{graphics}
\usepackage{xcolor}
\usepackage{grffile} 
\title{Monte Carlo N-Particle simulations of an underwater chemical threats detection system
using neutron activation analysis}
\author[a,1]{P.~Sibczy{\'n}ski,\note{Corresponding author.}}
\author[b]{M.~Silarski,}
\author[c]{O.~Bezshyyko,}
\author[b]{V.~Ivanyan,}
\author[b]{E.~Kubicz,}
\author[b]{Sz.~Nied{\'z}wiecki,}
\author[b]{P.~Moskal,}
\author[b]{J.~Raj,}
\author[b]{S.~Sharma,}
\author[c]{and O.~Trofimiuk}
\affiliation[a]{National Centre for Nuclear Research, Electronics and Detection Systems Division\\A. So{\l}tana 7, 05-400 Otwock, Poland}
\affiliation[b]{Jagiellonian University, Faculty of Physics, Astronomy and Applied Computer Science\\
S. {\L}ojasiewicza 11, 30-348 Krak{\'o}w, Poland}
\affiliation[c]{Taras Shevchenko National University of Kyiv\\Volodymyrska 60, 01033, Kyiv, Ukraine}
\emailAdd{pawel.sibczynski@ncbj.gov.pl}
\abstract{In this paper we present Monte Carlo N-Particle (MCNP) simulations of the system
for underwater threat detection using neutron activation analysis developed in the SABAT project.
The simulated system is based on a D-T neutron generator emitting 14~MeV neutrons without associated $\alpha$ particle detection and equipped with a LaBr$_3$:Ce scintillation detector offering superior energy resolution and allowing for precise identification of activation $\gamma$ quanta. The performed simulations show that using the neutron activation analysis method with the designed geometry we are able to identify $\gamma$-rays from hydrogen, carbon, sulphur and chlorine originating from mustard gas in a sea water environment. 
Our results show that the most efficient way of mustard gas detection is to compare the integral peak ratio for Cl and H.}
\begin{document}
\maketitle
\flushbottom
\section{Introduction}
\begin{figure}
	\centering
	\label{j._beldowski_et_al._chemsea_2015}
	\includegraphics[width=0.49\textwidth]{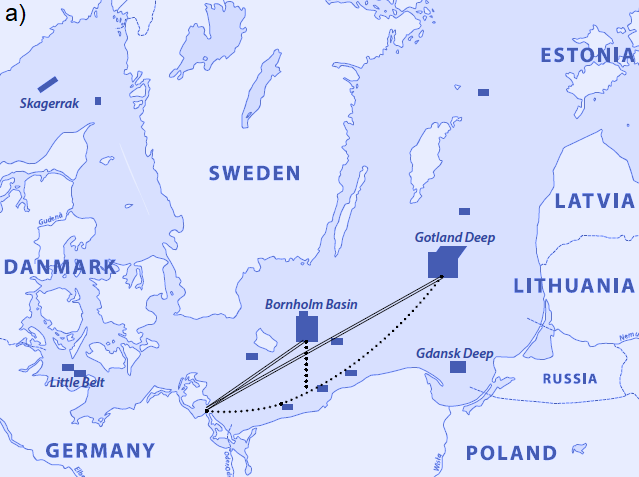}
	\includegraphics[width=0.50\textwidth]{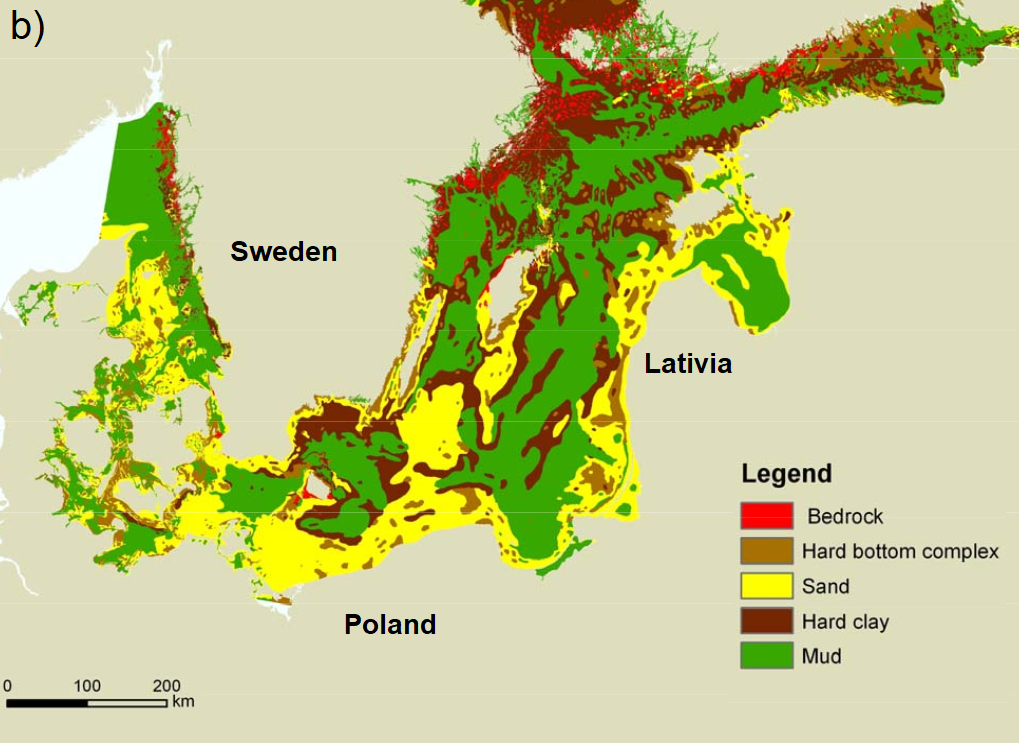}
	\caption{a) Map of the known areas of the dumped munition (blue rectangles).
	The solid and dashed curves show the official and unofficial transport routes,
	respectively. The figure is adapted from \cite{j._beldowski_et_al._chemsea_2015}.
	b) Map the Baltic Sea seabed composition, adapted from \cite{kyryliuk_total_2016}.}
\end{figure}

The SABAT project (Stoichiometry Analysis By Activation Techniques) aims to design and construct
a device for underwater threats detection using neutrons 
as a probe for nondestructive stoichiometry
determination~\cite{silarski_design_2016, silarski_device_2016}. The presently used methods are based
predominantly on sonars which provide only shapes of underwater objects and require additional inspection
of a diver or an underwater Remotely Operated Vehicle (ROV) to assess if the object is dangerous.
These methods are expensive, slow and do not provide identification of a substance inside the tested object.
New, more effective techniques are needed in particular for protection of harbors and off-shore infrastructure,
contraband uncovering and environmental protection, especially on sea areas of intensive warfare,
e.g. the Baltic Sea, where over 300 kilotons of munition was sunk. Depending on the source, from 40 to 65
kilotons of this arsenal are chemical agents \cite{andrulewicz_war_1996,
filipek_chemical_2014}. The main contaminated areas were determined within the CHEMSEA project
\cite{j._beldowski_et_al._chemsea_2015} and are presented in Fig.~\ref{j._beldowski_et_al._chemsea_2015}a.
Moreover, unknown amount of dangerous war remnants are spread over the whole Baltic, especially along
maritime convoys paths. Exceptionally dangerous is the chemical weapon dumped randomly near a seashore
since it can be thrown ashore during a storm. As it was shown in Fig.~\ref{j._beldowski_et_al._chemsea_2015}b
the munition is exposed to various environmental conditions due to different sediments composition
on the bottom of the Baltic Sea. This affects both the corrosion processes, detection possibility
and proper planning of removal of this ecological threat for which precise knowledge of the location
and amount of these hazardous substances is crucial.

Application of the Neutron Activation Analysis (NAA) may improve detection of underwater
dangerous substances providing non-destructive determination of their elemental content. 
It is based on neutron pulse emission
which excites nuclei of the investigated item. 
Detection of characteristic $\gamma$ quanta
produced in de-excitation of the nuclei then allows one to identify the
elemental composition of the substance~\cite{silarski_applications_2013, moskal_nuclear_2011}.
The NAA is widely used in explosives detection systems operating on the ground, such as {EURITRACK}
or {SWAN}~\cite{maglich_birth_2005, perot_euritrack_2006, kazmierczak_simple_2015}, but in the aquatic
environment one needs to overcome many difficulties related to technical aspects and strong background
radiation from oxygen and hydrogen. There is only one solution built and tested in the framework
of the UNCOSS (Underwater Coastal Sea Surveyor) project. However, it aimed only on the detection
of explosives~\cite{valkovic_underwater_2007, eleon_preliminary_2010}. An alternative solution
for a detector which uses NAA technique and 
special guides for neutrons and emitted $\gamma$-rays
was proposed by the SABAT group~\cite{silarski_device_2016}. The device allows for detection
of dangerous substances hidden deeper in the sea bottom with significantly reduced background and
decreased neutrons and $\gamma$-rays scattering. It also may provide determination of the density
distribution of the dangerous substance in the tested object
~\cite{silarski_applications_2013,silarski_project_2015,silarski_design_2016}.  

In this article we present feasibility studies of 
sulfide mustard (C$_4$H$_8$Cl$_2$S,
known also as a mustard gas) detection using the SABAT system without using the associated $\alpha$
particle measurement. The huge environmental background is reduced by the use of a detection trigger allowing for separate
measurements of delayed and prompt activation $\gamma$ quanta. For the latter, the use of neutron and gamma guide
tubes enhances significantly the signal to background ratio. 
\section{The MCNP simulations of the SABAT system performance}
\subsection{General model assumptions}
We have performed Monte Carlo simulations with the MCNP v6.11~\cite{goorley_initial_2012}
and MCNPX-POLIMI \cite{pozzi_mcnp-polimi:_2003} packages using the ENDF71x library \cite{conlin_continuous_2013}.
The high-performance notebook equipped with Core i7 processor and computing cluster at National Centre for Nuclear Research (NCBJ) {\'S}wierk Informatics Centre
(CI{\'S}) was used. As a commercially available neutron generator we simulated the operational conditions
and geometry of a Thermo Scientific P385 pulsed D-T neutron generator~\cite{ludewigt_neutron_2011}.
It emits 14~MeV neutrons and offers a neutron flux of 
3$\times$10$^8$ n/s with normal operation mode.
This neutron generator can emit neutrons in a very short pulse of about 5~$\mu$s with duty factor of 5\%.
In the UNCOSS project a $\diameter$3"~$\times$~3" LaBr:Ce $\gamma$-ray scintillation detector was proposed
for underwater explosives detection \cite{eleon_experimental_2011}, presenting excellent energy resolution
of 2.9\% at 662~keV $\gamma$-rays from $^{137}$Cs, high light yield and short decay
time~\cite{van_loef_high-energy-resolution_2001, van_loef_scintillation_2002, moszynski_study_2008, sibczynski_characterization_2017}. 
Following that recommendation, we included a $\diameter$ 2"$\times$2" LaBr$_3$:Ce scintillator into
the simulation code.\\
The 14~MeV neutrons, due to their high energy, excite nuclei predominantly in the process of inelastic
scattering which results in almost immediate emission of characteristic $\gamma$-rays. 
For a continuous
neutron beam some of the lines overlap with $\gamma$-quanta originating from the neutron capture 
which are emitted after neutrons thermalize. 
Thus, to separate the prompt $\gamma$-rays from the delayed ones we have simulated a short 5~$\mu$s neutron emission pulse, during which the data were acquired through 2~$\mu$s
in order to prevent registration of  rays originated from neutron capture. The detection of $\gamma$ radiation emitted from nuclei after neutron capture was simulated with the same neutron time emission of 5~$\mu$s as for prompt $\gamma$-rays detection, however, longer acquisition time window between 10~and~100~$\mu$s after the neutron generator pulse was assumed.
This time gate was chosen taking into account thermalization of the fast neutron in the container
with mustard gas, which can take from several $\mu$s up to even 100~$\mu$s.

\begin{figure}[!b]
	\centering
	\includegraphics[width=0.6\textwidth]{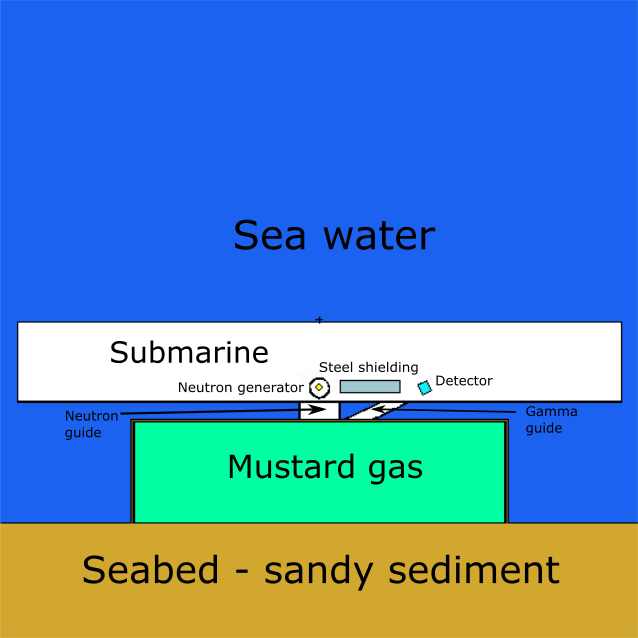}
	\caption{The simulated SABAT system geometry. A submarine (white rectangle)
	with the P385 D-T neutron generator (black circle) and $\gamma $quanta detector
	(light-blue) installed. The neutron and $\gamma$ quanta guide tubes
	(white, under the submarine) are simulated as cuboid and polyhedron, respectively.
	The irradiated mustard gas is placed on the bottom of the sea (green rectangle)
	in a 3~mm thick steel container. Detection of the mustard gas covered by 1~cm of wood and without wood was also evaluated. Details of the geometry can be found in the text.}
	\label{fig:15.0_geometry}
\end{figure}
The scheme of the simulated SABAT system geometry is presented in Fig. \ref{fig:15.0_geometry}.
A submarine (white rectangle) with dimensions 300~$\times$~300~$\times$~40~cm$^3$, made of
3~mm stainless steel, contains the 14~MeV neutron source (yellow dot) and cylindrical shape
$\diameter$2" $\times$ 2" LaBr$_3$:Ce $\gamma$-ray detector (light-blue) which is placed 50~cm from the target
in the neutron generator (cylinder around yellow dot). 
The neutron and $\gamma$ quanta guide tubes (white polyhedrons under the submarine)
are simulated as 20 $\times$ 20 $\times$ 10~cm$^3$ cuboid and 35~cm long polyhedron with a top and bottom base of 
20 $\times$ 20 cm$^2$ and 20 $\times$ 7.7~cm$^2$, respectively.
The $\gamma$ guide tubes are also made of 3~mm thick stainless steel. A container with mustard gas
with dimensions of 194 $\times$ 50 $\times$ 50~cm$^3$ in a 3~mm thick steel box (green rectangle)
is placed on the sea bottom represented by dark-yellow rectangle (400 $\times$ 400 $\times$ 100~cm$^3$). Two series of simulations were carried out - one with covering the mustard gas container with 1~cm of wood, and one without the covering. Both the submarine and guide tubes are filled with air under normal pressure. Materials composition, except mustard gas implemented according to the atomic fractions, were taken from the commonly available PNNL-15870 rev. 1 library~\cite{jr_compendium_nodate}. To evaluate performance of the detection system and estimate active background due to the presence of water,
we performed analogue simulations exchanging mustard gas with water.
The surrounding environment was simulated as water with 7.8\permil\ of salinity and traces of elements taken from \cite{kulik_physicochemical_1993} on page 12 in Table 1 on the last column, expressed in mg/kg. The values were normalized to water salinity. As a sea bottom, a sandy sediment was chosen, composed of 25\% of sea water and trace elements, presented in details in Table \ref{tab:sandy_sediment_composition}.
This conditions corresponds to the bottom of the Baltic Sea close to the shores
(to the depth of 10~m)~\cite{andrulewicz_chemical_2016}.
These areas are of the biggest interest since there one has 
no information about the amount of chemical
agents dumped into the sea (mainly because they were thrown out of ships in an uncontrolled way).
Moreover, the sandy bottom is common also in many other water reservoirs of intensive war operations.
According to literature the seashore close to Poland, including S{\l}upsk Furrow, is made
of mostly of sand~\cite{kyryliuk_total_2016}. 
We implemented this type of sediment in the simulation
model, as the presence of randomly dumped chemical weapon in this region is relatively high.
The sandy seabed contains much lower amount of heavy metals and is less saturated with water
than the seabed at the Gotland, Borholm and Gdańsk deep \cite{uscinowicz_trace_2011}. 
The contribution
of particulated organic compounds (POC) was introduced based on data provided
in~\cite{leipe_particulate_2011}.

	\begin{longtable}{|c|c|c|c|c|}
		\caption{Composition of sandy sediment used in the simulation model, according
		to~\cite{kyryliuk_total_2016, uscinowicz_trace_2011, leipe_particulate_2011, feistel_density_2010}.
		\label{tab:sandy_sediment_composition}}\\
		\hline
		Element 	& Mass ratio (\%) & Element & Mass ratio (\%) \\
		\hline
		H	        & 3.30144\%	      	&	Fe	&	1.00237\%	\\
		$^{10}$B			& 0.00001\%			&	Co	&	0.00050\%	\\
$^{11}$B	& 000002\%	&	Ni	&	0.00100\%	\\
C	& 0.35344\%	&	Cu	&	0.00100\%	\\
O	& 61.99990\%	&	Zn	&	0.00200\%	\\
F	& 0.00001\%	&	As	&	0.00080\%	\\
Na	& 1.34864\%	&	$^{79}$Br	&	0.00019\%	\\
Mg	& 0.00734\%	&	$^{81}$Br	&	0.00019\%	\\
Al	& 2.59482\%	&	$^{84}$Sr	&	0.00001\%	\\
Si	& 27.53650\%	&	$^{86}$Sr	&	0.00010\%	\\
S	& 0.00517\%	&	$^{87}$Sr	&	0.00007\%	\\
$^{35}$Cl & 0.08376\%	&	$^{88}$Sr	&	0.00086\%	\\
$^{37}$Cl & 0.02679\%	&	Cd	&	0.00010\%	\\
K	& 0.87891\%	&	Ba	&	0.00250\%	\\
Ca	& 0.84806\%	&	Hg	&	0.000002\%	\\
V	& 0.00100\%	&	Pb	&	0.00150\%	\\
Cr	& 0.00100\%	&		&  \\		
		\hline	
	\end{longtable}
%
Composition of the sea water
is based on the data from the PNNL database~\cite{jr_compendium_nodate} with salinity converted to
that present in the Baltic Sea (7.8\permil\ in the simulation model).
This is justified since, according to \cite{feistel_density_2010}, the ratio of elements traces
in ocean and Baltic Sea is the same in relation to the chlorine content.
The simulation method was based on the solution proposed by Evans~\cite{evans_monte_1998} with determination of the average flux of de-excitation $\gamma$-ray in the detector
	instead of calculation of particle flux through the detector surfaces. Finally, the MCNP simulation output data is normalized to one source neutron. Thus, it has to be normalized with the neutron source flux of a system and measurement time. Basics of the MCNP calculation are well reported in \cite{shultis_mcnp_2011}.
Since the detection method is based on separate acquisition of $\gamma$-rays originating from
inelastic neutron scattering and neutron capture we did separate simulations corresponding
to each case:
\begin{itemize}
	\item {for inelastic scattering we used F8 pulse height tally, with 2~$\mu$s neutron
	time cut-off and neutron full analog capture forcing. In fact, this kind of measurement
	should be done with lower neutron flux of about 10$^7$~n/s during 1000~s measurement time
	and 500 Hz pulse frequency in order to prevent a very high count rate in a very short time
	and significant number of pile-ups.}
\item{for neutron capture $\gamma$-rays detection the simulation was split into two stages: initially the F4 tally was used with the
T card for time gate measurement implementation. 
The output of this part of the simulation is an average flux of
$\gamma$-rays energy distribution. Then, as a second step, the F4 output was used for F8 photon pulse height tallying in the detector volume only, which result in obtaining the real shape of $\gamma$-ray energy spectra.
The detector physics - energy resolution, pair production
and peak Doppler broadening, was also introduced. For that simulation, we generated $3\times10^8$~n/s
during 100~s measurement time, which is achievable with the Thermo Scientific P385 neutron generator.}
\end{itemize}
%

\vspace{-0.4cm}
\begin{longtable}{|c|c|c|}
	\caption{The most prominent $\gamma$-rays emitted from the mustard gas and the surrounding
		environment after neutron activation. \label{tab:gamma_rays_mustgas}}\\
	\hline
	Energy (MeV) 	& Nucleus 	& Reaction type   \\
	\hline
	0.79  			& Cl		& Neutron capture\\
	1.17			& Cl		& Neutron capture\\
	1.78			& Si		& Inelastic scattering\\
	1.94			& Cl		& Neutron capture\\
	2.12			& Cl		& Inelastic scattering\\
	2.23			& H			& Neutron capture\\
	2.23			& S			& Inelastic scattering\\
	4.44			& C			& Inelastic scattering\\
	6.12      		& Cl    & Neutron capture\\
	6.13      		& O     & Inelastic scattering\\
	7.64			& Fe    & Neutron capture\\
	7.79			& Cl		& Neutron capture\\
	8.58     		 & Cl    & Neutron capture\\
	9.30			& Fe	  & Neutron capture\\
	\hline	
\end{longtable}
The pulse height tally was used with the energy threshold set to 100~keV. The energy resolution
of the simulated LaBr$_3$:Ce detector was included by the following full-width-at-half-maximum
(FWHM) parametrization set with the GEB card~\cite{sibczynski_characterization_2017}:
FWHM$(E_{\gamma}) = a + b\sqrt{E_{\gamma} + cE_{\gamma}^{2}}$, where $E_{\gamma}$ denotes the
$\gamma$-quantum energy (in MeV), $a=2.0\cdot 10^{-4}$~MeV, $b=2.2\cdot 10^{-2}$~MeV$^{1/2}$
and $c=0.5$~MeV$^{-1}$. This parametrization gives the FWHM of about 2\% for 4.44~MeV $\gamma$-rays. The generated raw energy spectra with 10~keV energy bins were normalized to the number of histories which were set to 10$^{10}$ for prompt $\gamma$-rays detection and 10$^9$ for neutron capture $\gamma$-rays.\\
The most intense $\gamma$-rays emitted from materials used in the simulations are summarized
in Tab.~\ref{tab:gamma_rays_mustgas}. To demonstrate the feasibility of mustard gas detection
with SABAT we focused on 2.12~MeV and 6.12~MeV lines of chlorine, 2.23~MeV peak of sulphur
and 4.4~MeV $\gamma$ quanta from carbon. The presence of mustard gas can be detected by analysis
of peak ratios for the mentioned lines and the 6.13~MeV chlorine and 2.23~MeV hydrogen peaks
(for prompt and delayed $\gamma$-rays, respectively). We have determined these ratios for simulations
done with and without neutron and $\gamma$ quanta guide tubes and for different distances between the
submarine and mustard gas container.
\subsection{Mustard gas detection with de-excitation $\gamma$-rays}
\begin{figure}[!b]
	\centering
	\includegraphics[width=0.55\textwidth]{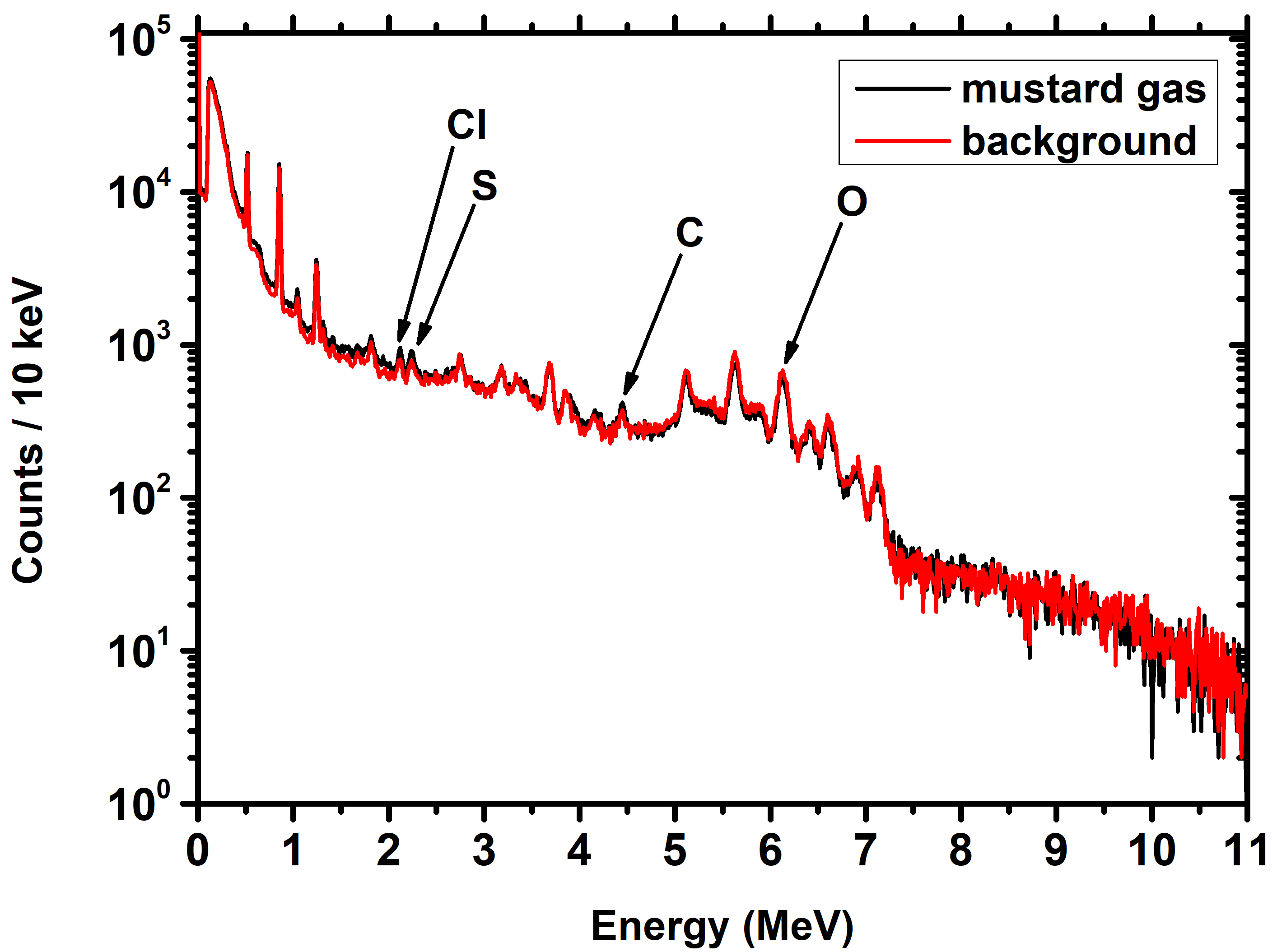}
	\includegraphics[width=0.55\textwidth]{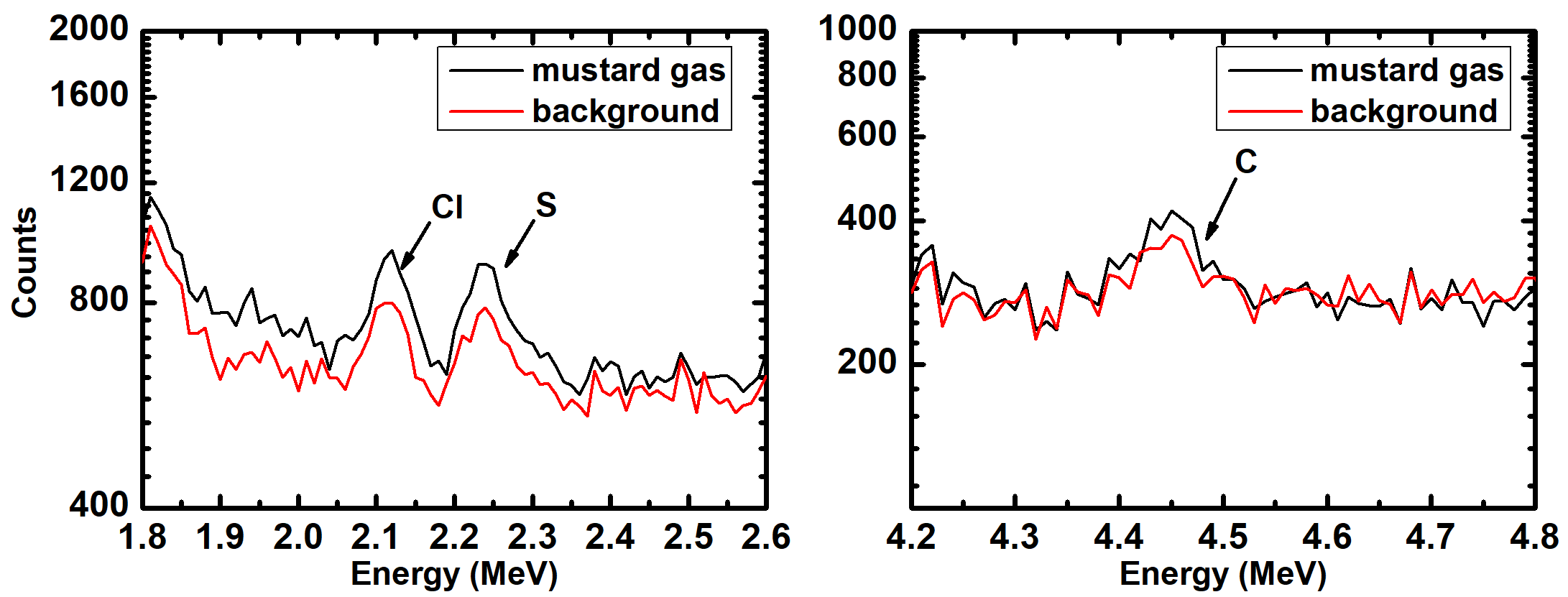}
	\caption{The simulated energy spectra of prompt $\gamma$-rays registered with the
		LaBr$_3$:Ce scintillator for the case of mustard gas presence (black) and for background
		(red). In the lower panel zooms of the energy spectra around the chlorine, sulphur and carbon
		peaks are presented. The inspection system was equipped with $\gamma$ and neutron guide tubes.}
	\label{fig:v17.0-v17.1_graph_prompt_spectra}
\end{figure}

Distribution of the prompt $\gamma$-rays energies obtained with the sensor equipped
in gamma guide tubes and positioned 10~cm above the mustard gas is shown in
Fig.~\ref{fig:v17.0-v17.1_graph_prompt_spectra}. 

One can see three $\gamma$ lines related to
the neutrons inelastic scattering on mustard gas, 2.23~MeV from sulphur, 4.44~MeV from carbon
and a weak 2.12~MeV line of chlorine which is barely visible and overlaps with the $\gamma$-ray peak from iron.
Thus, its detection is notably limited. 
The line from sulphur covers ideally with the energy
of $\gamma$-rays from hydrogen (2.23~MeV).
However, the latter are emitted after the thermal
neutron capture, a long time after neutron pulse.\\
The estimated count rate using the acquisition time gate of 2~$\mu$s in the presence of mustard gas
is about 1500 counts per second (cps) assuming neutron flux of $10^7$~n/s, 500 Hz beam pulse frequency energy threshold of 100~keV and time acquisition gate of 2$\mu$s. It is important to emphasize that due to a very short neutron emission period, as well as the acquisition time, further increasing of the neutron flux could create significant
amount of pile-ups in the scintillation detector and decrease the detection performance.
\subsection{Mustard gas signature with delayed $\gamma$-rays detection}
In the previous section we showed the feasibility of mustard gas detection with prompt $\gamma$-rays.
Alternatively, detection of radiation originating from neutron capture could be applied for
interrogation of objects in the aquatic environment. According to the MCNP simulations, this method
is much more efficient than the method based on prompt $\gamma$-rays from neutron inelastic scattering for detection of mustard gas.
\begin{figure}[!b]
	\centering
	\includegraphics[width=0.5\textwidth]{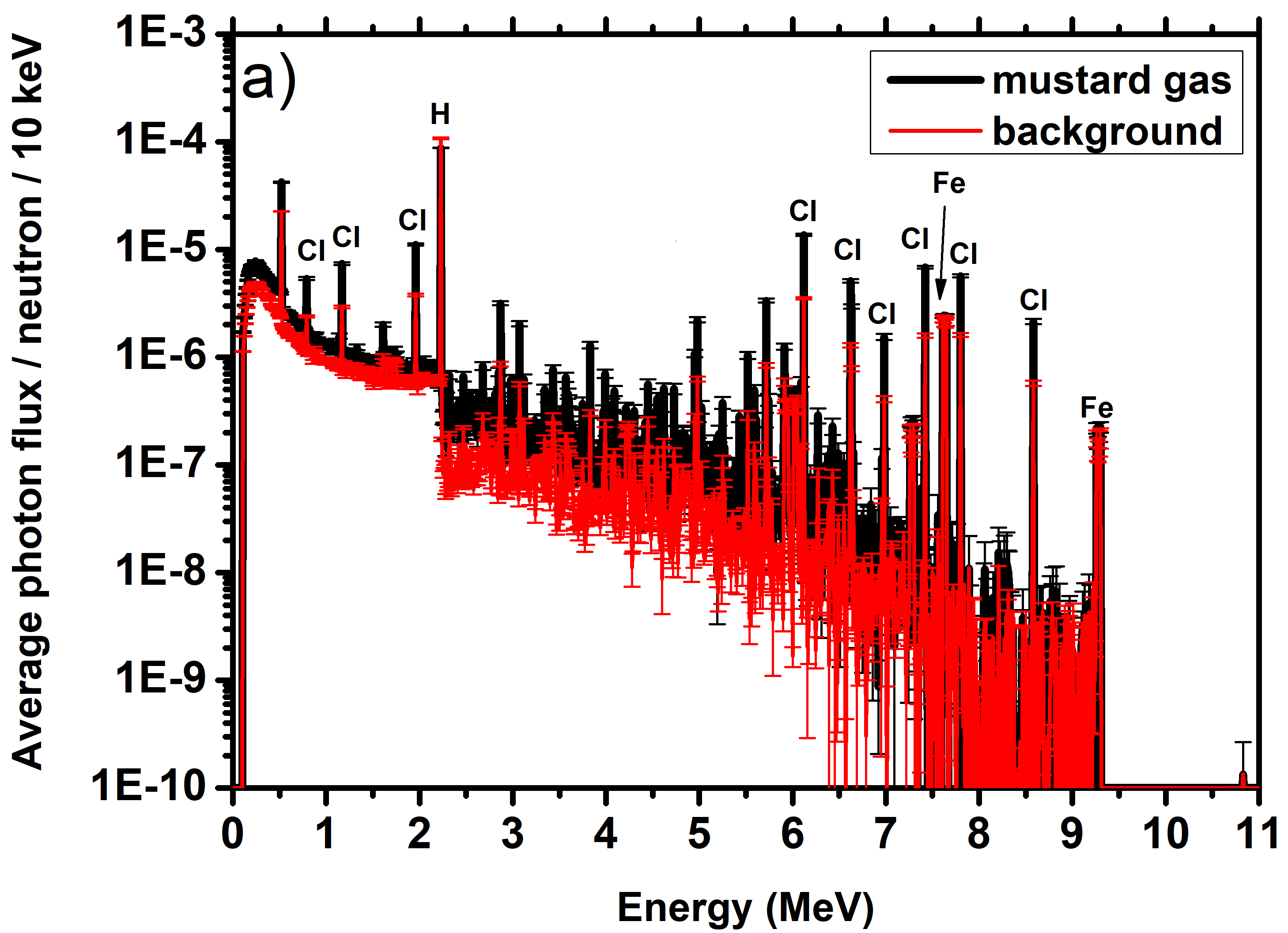}
	\includegraphics[width=0.49\textwidth]{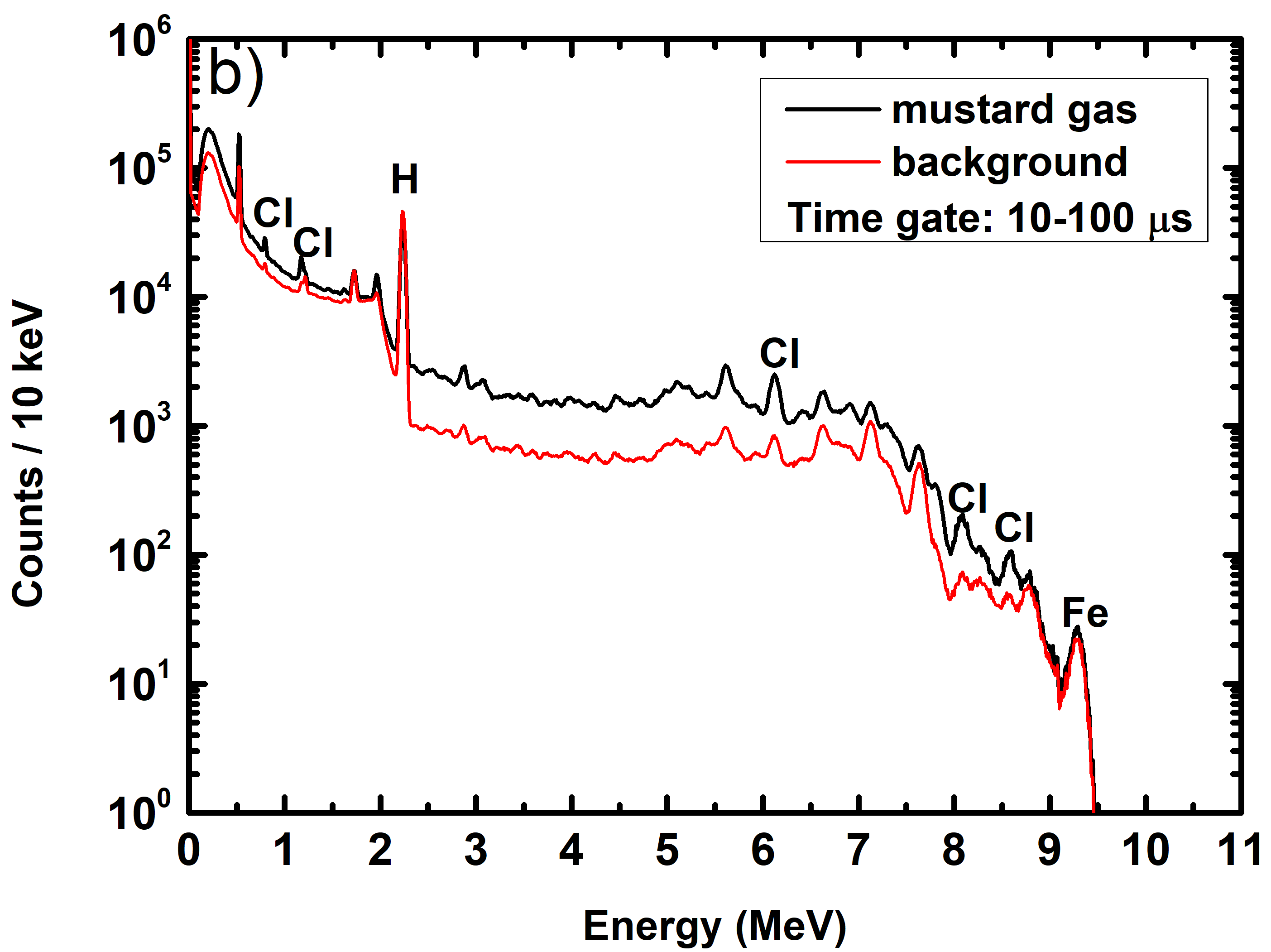}
	\caption{Delayed $\gamma$-rays average flux distribution (a) and energy spectrum (b) in case of
	mustard gas presence (black) and for background only (red). The acquisition time gate after neutron
	emission was set to 10~-~100 $\mu$s, with 5~$\mu$s long neutron emission pulse. The total simulated
	statistics corresponds to 100~s of measurement with flux of $3\times10^8$ n/s.
	The inspection system was equipped with $\gamma$ quanta and neutron guide tubes.}
	\label{fig:v11.0_v11.11_5-1000us_raw_spectra_v3}
\end{figure}
In Fig.~\ref{fig:v11.0_v11.11_5-1000us_raw_spectra_v3}a an average photon flux distribution for
mustard gas and background is presented. The inspection system was again equipped with $\gamma$ quanta
and neutron guide tubes. The significant excess of counts from Cl for mustard gas is clearly seen.
We also see $\gamma$-rays from $^{54}$Fe at 9.3~MeV as well as peak at 7.64~MeV from $^{56}$Fe.
The main peaks of interest from Cl are 1.17~MeV, 6.12~MeV, 7.79~MeV and 8.58~MeV.
Although a peak from oxygen is known to be present at 6.13~MeV it does not overlap with the chlorine.
This is due to the fact that the oxygen peak originates from the fast neutron inelastic scattering.
The energy distribution of the neutron capture $\gamma$ quanta is presented in
Fig.~\ref{fig:v11.0_v11.11_5-1000us_raw_spectra_v3}b. The Cl peak at 6.12~MeV is well visible.

\begin{figure}[!t]
	\centering
	\includegraphics[width=0.62\textwidth]{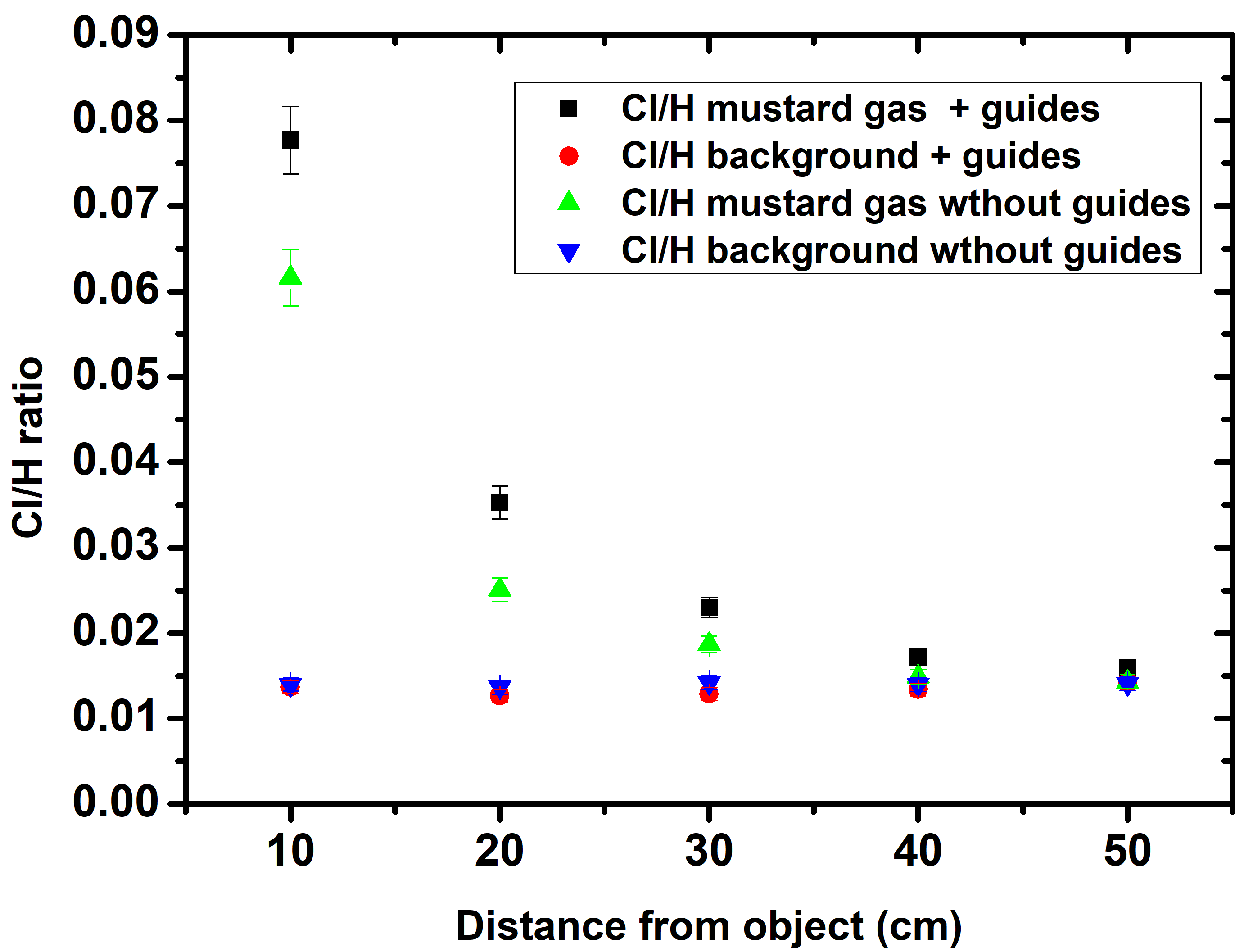}
	\caption{The obtained Cl/H ratio as a function of distance between the mustard gas container
		and the SABAT sensor simulated for mustard gas presence and background only.}
	\label{fig:Cl_H_ratio_vs_distance}
\end{figure}

\subsection{Elemental peak ratios as signatures of the simulated chemical agent}
As it was mentioned before, the detection of mustard gas can be performed by analysis of
ratios of peaks for chlorine, sulphur and carbon to the oxygen peak (prompt $\gamma$-rays)
and to the hydrogen line (delayed quanta). The peak ratios for all the elements of interest with the sensor distant by 10~cm from the mustard gas container are presented in Tab.~\ref{tab:stoichiometry_mustgas}.

\begin{longtable}{|c|c|c|c|c|}
	\caption{Peak ratios obtained for mustard gas and background with the simulated detection
		system equipped with $\gamma$ and neutron guides and without them. Results are presented for
		the submarine positioned 10~cm above the threat.
		\label{tab:stoichiometry_mustgas}}\\
	\hline
	\multicolumn{5}{|c|} {Inelastic scattering prompt $\gamma$-rays - system with gamma guides}\\
	\hline
	Peak ratio 	& Energy (MeV)	& 	Threat 	& Mustard gas & Background\\
	\hline	
	Cl/O 		& 2.12/6.13	&	Mustard gas & 0.37 $\pm$ 0.04	& 0.22 $\pm$ 0.03\\
	S/O 		& 2.23/6.13	&	Mustard gas & 0.41 $\pm$ 0.04	& 0.23 $\pm$ 0.03\\
	C/O 		& 4.44/6.13	&	Mustard gas & 0.24 $\pm$ 0.03	& 0.13 $\pm$ 0.02\\
	\hline
	\multicolumn{5}{|c|} {Neutron capture $\gamma$-rays - system with gamma guides}\\
	\hline	
	Cl/H & 6.12/2.23	&	Mustard gas & 0.078 $\pm$ 0.003	& 0.014 $\pm$ 0.002\\
	\hline\hline
	\multicolumn{5}{|c|} {Inelastic scattering prompt $\gamma$-rays - system without gamma guides}\\
	\hline
	Peak ratio 	& Energy (MeV)	& 	Threat 	& Mustard gas & Background\\
	\hline	
	Cl/O 		& 2.12/6.13	&	Mustard gas & 0.23 $\pm$ 0.02	& 0.17 $\pm$ 0.02\\
	S/O 		& 2.23/6.13	&	Mustard gas & 0.27 $\pm$ 0.03	& 0.19 $\pm$ 0.02\\
	C/O 		& 4.44/6.13	&	Mustard gas & 0.20 $\pm$ 0.02   & 0.13 $\pm$ 0.02\\
	\hline
	\multicolumn{5}{|c|} {Neutron capture $\gamma$-rays - system without gamma guides}\\
	\hline	
	Cl/H & 6.12/2.23	&	Mustard gas & 0.062 $\pm$ 0.002	& 0.014 $\pm$ 0.002 \\
	\hline
\end{longtable}

\newpage

For the prompt $\gamma$-rays only the Cl/O and S/O ratio is greater for mustard gas than for the background by about $3\sigma$. On the other hand, detection of the delayed $\gamma$-rays provides much better
sensitivity with the Cl/H ratio.
With the P385 neutron generator optimized for the delayed $\gamma$-ray detection this ratio for simulated mustard gas with use of neutron and gamma guide tubes is
equal to 0.078 $\pm$ 0.003, comparing to the same value for background: 0.014 $\pm$ 0.002.
Since we expect a dependence of performance of the detector on the distance to the inspected object
the simulations were done for different positions of the device.
In Fig.~\ref{fig:Cl_H_ratio_vs_distance} we summarize the Cl/H ratios as a function of the distance
of the SABAT sensor and the mustard gas for system with and without guide tubes.

In turn the ratio in case of the mustard gas presence change with the distance and is
much higher for the device equipped with the guide tubes. According to the simulations the sensitivity of
the SABAT system elevated 50~cm above the inspected object is so low that one cannot see any signal
of the mustard gas presence.\newline
\indent According to literature \cite{szarejko_baltic_2009}, some of the dumped chemicals were placed in wooden boxes. We evaluated the case when the mustard gas inside metal container is covered by 1 cm of wood. The wood composition was taken from PNNL library \cite{jr_compendium_nodate}. However, we modified the wood density to 1.2 g/cm$^3$ and set 50\% of its composition as the sea water, as it is strongly absorbed in wood. Simulations were done for the SABAT device positioned 10 cm above the inspected object. Results show that detection sensitivity with use of Cl/H ratio measurement method for that case is the same in comparison with the model without the wooden box. Concerning the prompt $\gamma$-rays detection, the slight signal above the background level was observed. Results were summarized in Table \ref{tab:stoichiometry_mustgas_wood}.

\begin{longtable}{|c|c|c|c|c|}
	\caption{Peak ratios obtained for mustard gas covered with 1 cm of wood and background with the simulated detection system equipped with $\gamma$ and neutron guides and without them. Results are presented for the submarine positioned 10~cm above the threat.
		\label{tab:stoichiometry_mustgas_wood}}\\
	\hline
	\multicolumn{5}{|c|} {Inelastic scattering prompt $\gamma$-rays - system with gamma guides}\\
	\hline
	Peak ratio 	& Energy (MeV)	& 	Threat 	& Mustard gas & Background\\
	\hline	
	Cl/O 		& 2.12/6.13	&	Mustard gas &  0.31 $\pm$ 0.03	& 0.22 $\pm$ 0.03\\
	S/O 		& 2.23/6.13	&	Mustard gas &  0.39 $\pm$ 0.04	& 0.23 $\pm$ 0.03\\
	C/O 		& 4.44/6.13	&	Mustard gas &  0.27 $\pm$ 0.03	& 0.13 $\pm$ 0.02\\
	\hline
	\multicolumn{5}{|c|} {Neutron capture $\gamma$-rays - system with gamma guides}\\
	\hline	
	Cl/H & 6.12/2.23	&	Mustard gas & 0.077 $\pm$ 0.003	& 0.014 $\pm$ 0.002\\
	\hline\hline
	\multicolumn{5}{|c|} {Inelastic scattering prompt $\gamma$-rays - system without gamma guides}\\
	\hline
	Peak ratio 	& Energy (MeV)	& 	Threat 	& Mustard gas & Background\\
	\hline	
	Cl/O 		& 2.12/6.13	&	Mustard gas &  0.24 $\pm$ 0.02	& 0.17 $\pm$ 0.02\\
	S/O 		& 2.23/6.13	&	Mustard gas &  0.27 $\pm$ 0.03	& 0.19 $\pm$ 0.02\\
	C/O 		& 4.44/6.13	&	Mustard gas &  0.21 $\pm$ 0.03  & 0.13 $\pm$ 0.02\\
	\hline
	\multicolumn{5}{|c|} {Neutron capture $\gamma$-rays - system without gamma guides}\\
	\hline	
	Cl/H & 6.12/2.23	&	Mustard gas & 0.062 $\pm$ 0.002	& 0.014 $\pm$ 0.002 \\
	\hline
\end{longtable}

\section{Summary}
The first feasibility studies of the SABAT detection system revealed that the mustard gas detection
is possible in the aquatic environment even without the associated $\alpha$ particle measurement.
Based on the performed MCNP simulations we conclude that the separate detection of prompt $\gamma$-quanta
and radiation from the delayed thermal neutron capture gives the best performance of the detection system.
For the prompt $\gamma$-rays signal for sulphur and chlorine can be observed if the LaBr$_3$:Ce detector
is used, although the sensitivity for the prompt chlorine $\gamma$-rays is rather low.
This measurement requires a neutron generator allowing 
for low repetition time between 100 - 1000 Hz and low duty cycle of about 5\% 
to suppress $\gamma$-rays originating from thermal neutron capture.
Otherwise the signal from sulphur could be masked by a significant number of counts from the hydrogen line of exactly the same energy. 
The acquisition system used for prompt $\gamma$-rays detection should
be able to operate at high count rates in a short time period,
up to 1~Mcps~\cite{korolczuk_digital_2016,korcyl_sampling_2016,korcyl_evaluation_2018}.\\
The simulations for $\gamma$-quanta originating from the neutron capture clearly show
the feasibility of chlorine detection, even if it is present in the sea water.
This measurement appears to be much more efficient than the detection of prompt $\gamma$-rays. 
The use of $\gamma$ and neutron guide tubes improve performance of the delayed radiation detection by 20$\%$ - 40$\%$,
depending on the distance from the object which should not be greater than 30~cm in order to achieve
sufficient sensitivity.
The inelastic scattering prompt $\gamma$-rays detection can be considered as a supplementary analysis,
providing additional information after chlorine. It should be stressed that the detection of sulphur presence may be important,
as the mustard gas can be covered by organic compounds, which after exposition to neutrons emit 4.44~MeV
$\gamma$-rays from carbon.\\
The possibility of successful mustard gas detection using neutron generator without the associated
particle measurement allows for significant reduction of the cost and complexity of the device. However,
the recoil $\alpha$ particle registration would not only reduce further the background, but may also
give opportunity to obtain tomographic-like image of the inspected object. 
Knowing position of the $\alpha$ particle interaction in the generator and the location of the
$\gamma$-rays hit in the detector one can use the time difference between these two measurements to determine
the point of emission of the $\gamma$ quantum (detailed explanation of this idea can be found
in~\cite{silarski_project_2015}). Simulations show that the elemental ratios for materials which can be likely found on the bottom of the Baltic Sea are much different from the one for mustard gas (e.g. the Cl/H ratio for a wooden box was found to be 0.014 $\pm$ 0.002). Steel elements, which are often found on the Baltic seabed, can be also discriminated due to lack of Cl and H in their composition, thus, would not affect to Cl/H ratio. This shows that SABAT sensor will be able to discriminate real threats from the safe items on the bottom of the sea. Thus, our future studies will focus on checking feasibility of other substances detection (e.g. Clark I or Clark II gases) and possible improvements given by
the associated particle measurement.
\acknowledgments
We would like to thank prof. Steven Bass for proofreading of the article and many useful comments.
We acknowledge support from the Polish National Centre for Research and Development through
grant No. LIDER/17/0046/L-7/15/NCBR/2016.
\bibliographystyle{JHEP}
\bibliography{sbt_bib_v8}
\end{document}